\documentclass[final]{aipproc}
\layoutstyle{6x9}

\SetInternalRegister\hbadness{8000} % pseudo latin isn't breaking very well :-)

\begin{document}

\title[]{Phase transitions and  behavior of kaons   in hot and  dense matter}

\author{C. A. de Sousa}{address={Centro de F\'{\i}sica Te\'{o}rica,
Departamento de F\'{\i}sica, Universidade, P3004-516 Coimbra, Portugal}}
%\email{celia@teor.fis.uc.pt}

\iftrue
\author{P. Costa}{address={Centro de F\'{\i}sica Te\'{o}rica,
Departamento de F\'{\i}sica, Universidade, P3004-516 Coimbra, Portugal}}
%\footnote{pcosta@fteor5.fis.uc.pt}

\iftrue
\author{M. C. Ruivo}{address={Centro de F\'{\i}sica Te\'{o}rica,
Departamento de F\'{\i}sica, Universidade, P3004-516 Coimbra, Portugal}}
%\footnote{maria@teor.fis.uc.pt}

\author{Yu. L. Kalinovsky}{address={Universit\'{e} de Li\`{e}ge, D\'{e}partment de Physique B5, Sart Tilman, B-4000, LIEGE 1, Belgium}}
%\footnote{kalinov@nusum.jinr.ru}
\fi

%%%%%%%%%%%%%%%%%%%%%%%%%%%%%%%%%%%%%%%%%%%%%%%%%%%%%%%%%%%%%%%%%%%%%%%%%%%%%%%%%%%%%%%%%%%%%%%%%%%%%%%%%%%%%%%%%

\begin{abstract}
We study phase transitions and the behavior of kaons in hot and dense matter, giving special attention to the role of strange quarks. At $T=0$, it is found that the fraction of the strange valence quarks  affects the energy per particle of the system, without changing the nature of the phase transition, and, on the other side, it has meaningful effects on the  behavior of kaons and antikaons masses.
The phase behavior of kaons in the $T-\rho$ plane is analyzed in connection with the chiral phase transition. 
\end{abstract}
\maketitle

%%%%%%%%%%%%%%%%%%%%%%%%%%%%%%%%%%%%%%%%%%%%%%%%%%%%%%%%%%%%%%%

Experimental and theoretical efforts have been done  in order to explore the $\mu-T$ phase boundary.
There are indications from Lattice QCD that the transition from the hadronic phase to the quark gluon plasma   occurs at temperatures of about 150 MeV. 
While  the phase transition with finite chemical potential and zero temperature is expected to be first order, at zero chemical potential and finite temperature there will be a smooth crossover.

Strange quark matter (SQM) has attracted a lot of interest since  the suggestion  that it could be the absolute ground state of matter. Stable SQM in $\beta$--equilibrium is expected to exist in the interior of neutron stars or might  also be formed in the earlier stages of heavy-ion collisions (in this case $\beta$--equilibrium may not be achieved). 

In this paper, we start with the  study of the stability condition and order of phase transition at $T=0$ and  $T\neq 0$. Several strange quark matter scenarios are considered, with and without $\beta$-equilibrium.  

Our research was carried out in the framework of the SU(3) Nambu-Jona-Lasinio (NJL) model with 't Hooft interaction. The Lagrangian is thus given by \cite{klev2}:
\begin{eqnarray}
{\mathcal L} &=& \bar{q} \left( i \partial \cdot \gamma - \hat{m} \right) q
+ \frac{g_S}{2} \sum_{a=0}^{8}
\Bigl[ \left( \bar{q} \lambda^a q \right)^2+
\left( \bar{q} (i \gamma_5)\lambda^a q \right)^2
 \Bigr] \nonumber \\
&+& g_D \Bigl[ \mbox{det}\bigl[ \bar{q} (1+\gamma_5) q \bigr]
  +  \mbox{det}\bigl[ \bar{q} (1-\gamma_5) q \bigr]\Bigr] \,.\\ \nonumber
\end{eqnarray}

Following a standard hadronization procedure and the  well known Matsubara technique, an effective action is obtained, allowing to calculate several physical quantities  at finite temperature and density \cite{costaruivo}.

{\bf Chiral phase transition at zero temperature}. We start by analyzing the behavior of quark matter at zero temperature and, in order to discuss the role of the strangeness degree of freedom, we will consider "neutron" matter in chemical equilibrium and with charge neutrality, and matter  without $\beta$--equilibrium \cite{costaruivo,CRSKprc}.
For matter without $\beta$-equilibrium we consider three cases: Case I -- "neutron" matter without strangeness, ($\rho_d=2 \rho_u\,\,, \rho_s=0$); Case II -- matter with equal chemical potentials ($\mu=\mu_d=\mu_u=\mu_s$) with isospin symmetry, $\rho_u=\rho_d,\,\rho_s=\frac{1}{\pi^{2}}(\mu^{2}-M_{s}^{2})^{3/2}\theta(
\mu^{2}-M_{s}^{2})$; Case III --   matter entirely flavor symmetric ($\rho_d=\rho_u=\rho_s$). 
We observe that  the pressure has a zero at $\rho_n=0$, the energy per baryon being $3\,M_u$ for non strange quark matter.
If there is another zero of the pressure, at $\rho_n\not=0$, that corresponds to a minimum of the energy, the criterion for stability of the system at that point is $\mu_u+2\mu_d < 3 M_u$ \cite {buballa,CRSKprc}.

We notice that in all cases, whether valence strange quarks are present or not, there is an absolute minimum of the energy per particle at non zero density and zero pressure, lower than  the vacuum constituent quark masses; so we have a first order phase transition with the formation of quark droplets (see Fig. 1, left side). 

{\bf Chiral phase transition at finite temperature and density}. Now we discuss the phase transition in hot and dense matter and we consider only matter in $\beta $--equilibrium (the other cases are qualitatively similar). We observe that for very low temperatures the absolute minimum of the energy turns to be at zero density: the phase transition is still first order but the system is unstable against expansion. With increasing temperature, we will have a crossover above $T > T_{cl}= 56$ MeV. The critical end point, that connects the first order phase transition and the crossover  regions, is found  at $T= 56\mbox{ MeV}$ and $\rho_n=1.53 \rho_0$.
The numerical results indicate   a clear manifestation of the restoration of chiral symmetry for the light quarks  with increasing temperature and density. A more smooth behavior of the strange quark is observed, with  the  chiral symmetry showing a slow tendency to get restored in the strange sector.
%%%%%%%%%%%%%%%%%%%%%%%%%%%%%%%%%%%%
\begin{figure}[t]
\vspace{-2cm}
\hspace{0.35cm}\includegraphics[width=6.75cm,height=6.75cm]{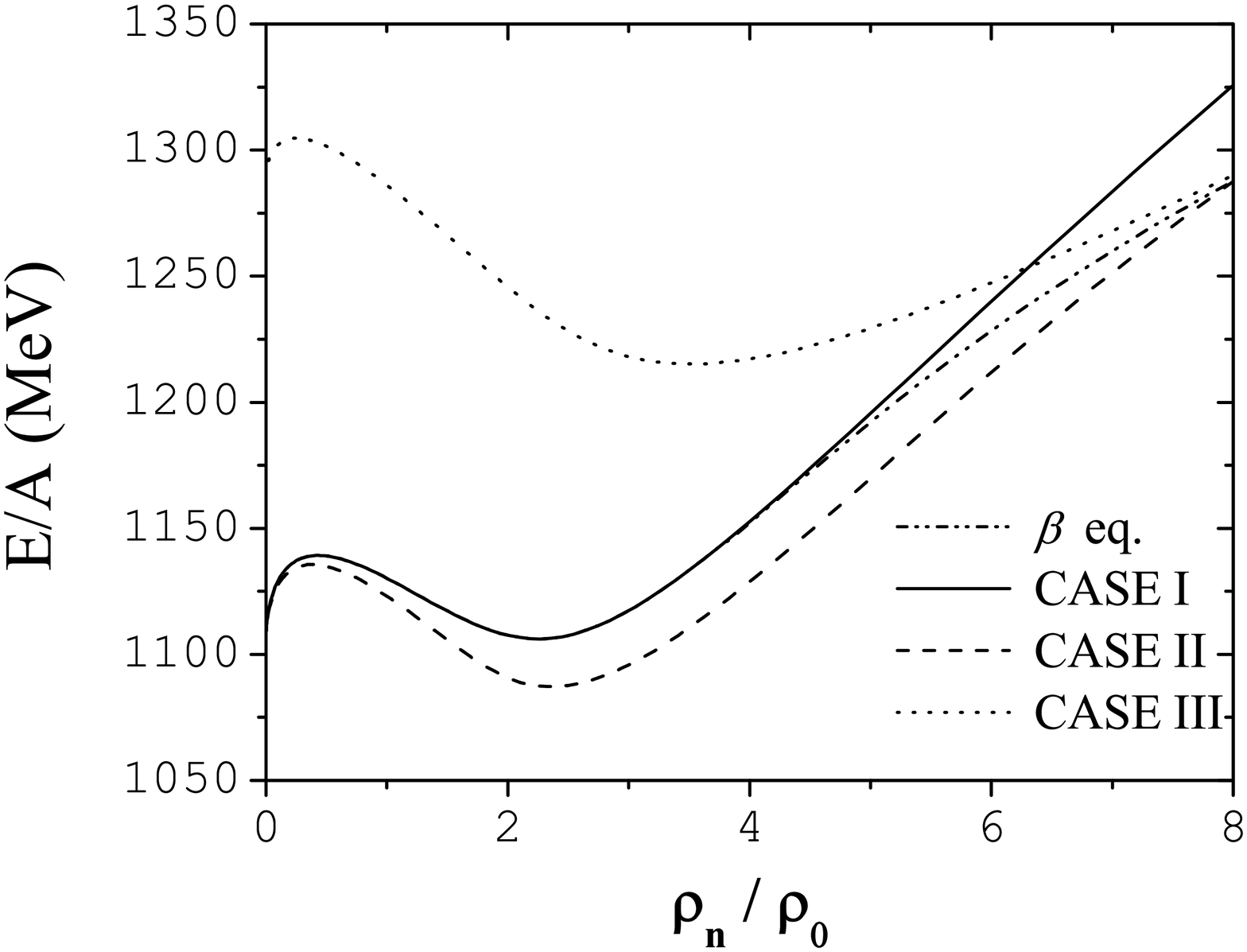}
\hspace{-0.85cm}\includegraphics[width=9.7cm,height=6.5cm]{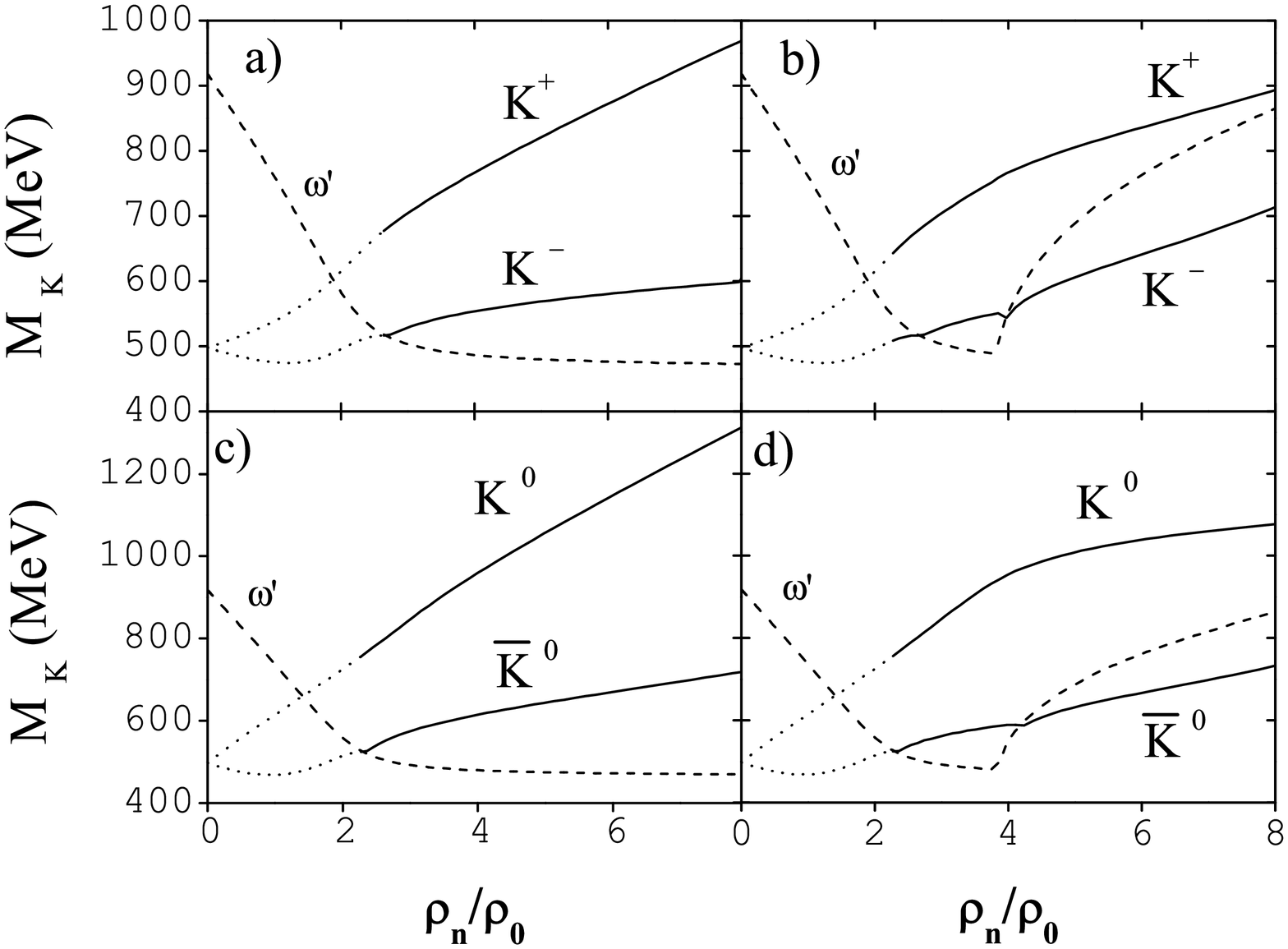}
\caption{Energy per baryon number for all types of quark matter considered at $T=0$ (left side). Kaon and anti-kaon masses as function of density at $T=0$ (right side):
Case I (a) and c)) and  with $\beta$--equilibrium (b) and d)). $\omega$' -lower limit of the Dirac sea continuum.}
\end{figure}
%%%%%%%%%%%%%%%%%%%%%%%%%%%%%%%%%%%%
%
{\bf Kaons in cold quark  matter}.
As it was already shown in other works \cite{costaruivo}, two kinds of solutions may be found in asymmetric matter for kaonic modes in the NJL model,  corresponding respectively to excitations of the Dirac sea and excitations of the Fermi sea; here, only  excitations of the Dirac sea will be discussed. In order to appreciate the role of the strangeness degree of freedom, we start by comparing the results obtained in matter with $\beta$--equilibrium and without, Case I.
We observe the expected splitting between charge multiplets: the increase of the masses of $K^+ ,K^0$  with respect to those of $K^-, \bar K^0 $, respectively, is due to Pauli blocking, and at the critical density the antikaons enter in the continuum.
By comparing the kaonic behavior in matter with  and without strange quarks (See Fig. 1, right side) one concludes that the presence of strange quarks has two effects: it is responsible for the antikaons becoming again bound states above a certain density ($\approx4\rho_0$) and contributes to reduce the splitting between kaon and antikaon masses.

{\bf Kaons in hot and dense matter}. 
In order to illustrate the combined effect of temperature and density on the behavior of kaons, we discuss the case of hot matter in weak equilibrium. It is found that, since the threshold of the $\bar q q$ continuum is now at the sum of the constituent quark masses, the mesons dissociate at densities and temperatures close to the critical ones.
The numerical results indicate a line (a kind of Mott circle)  separating the region where the meson are bound states from the region where they are in the continuum. 

To conclude and summarize, in the present paper we have investigated phase transitions in hot and dense matter, and the  in--medium behavior of kaonic mesons.
For a suitable choice of the parameters at zero temperature we have a mixed phase.  
We notice that, in flavor asymmetric matter, the minimum of the energy  is in a region where strange valence quarks are still absent. However, for higher densities the energy density is reduced by having three Fermi seas instead of just two.  Only for the case of equal number of quarks $u\,,d\,,s$ we found stable SQM, but with a higher energy per particle then atomic nuclei.  
Concerning the masses of the kaons, there is a splitting between the flavor
multiplets in flavor asymmetric matter.
In the high density region  the splitting is reduced in matter with strangeness.
In  hot and dense matter the phase transition becomes a crossover above the critical end point, $T= 56$ MeV and $\rho=1.53 \rho_0$, the system having a mixed phase before that point.  
The main feature is the dissociation of mesons at the Mott transition point that occurs when the meson masses equals the sum of the masses of their constituents. After that point the mesons cease to be bound states and become resonances. 

%%%%%%%%%%%%%%%%%%%%%%%%
\vspace{0.5cm}
Work supported by grant SFRH/BD/3296/2000 (P. Costa), CFT and by FEDER/FCT under project POCTI/FIS/451/94. 
%%%%%%%%%%%%%%%%%%%%%%%%

%%%%%%%%%%%%%%%%%%%%%%%%%%%%%%%%%%%%%%%%%%%%%%%%%%%%%%%%%%%%%%%%%%%%%%%%%%%%%%%%%%%%%%%%%%%%%%%%%%%%%%%%%%%%%%%%%%%%%%%%%%%%%%%%%%%%%%%%%%%%%%%%%%%%%%%%%%


\begin{thebibliography}{99}

\bibitem{klev2}
			P. Rehberg, S. P. Klevansky and J. H\"{u}fner, 
			{\it Phys. Rev.}  {\bf C 53}, 410 (1996).


\bibitem{costaruivo} 
			P. Costa and M. C. Ruivo, 
			{\it Europhys. Lett.} {\bf 60}(3), 356 (2002); 
			P. Costa, M. C. Ruivo and Yu. L. Kalinovsky, 
			{\it Phys. Lett.} {\bf B 560}, 171 (2003);
			{\it Phys. Lett.} {\bf B 577}, 129 (2003).
\bibitem{buballa}  
			M. Buballa and M. Oertel, 
			{\it Nucl. Phys.} {\bf A 642}, 39 (1998);
      {\it Phys. Lett.} {\bf B 457}, 261 (1999).
      
\bibitem{CRSKprc} 
			P. Costa, M. C. Ruivo, C. A de Sousa and Yu. L. Kalinovsky, 
			{\it Phys. Rev.}  {\bf C 70}, 025204 (2004).

\end{thebibliography}
\end{document}